\documentclass[amssymb,12pt,showpacs,aps,nofootinbib,showkeys]{revtex4}
\usepackage{graphicx}
\begin{document}
\title{Topological Maps from Signals}
\date{\today}
\author{Yu. Dabaghian$^1$, A. G. Cohn$^2$, L. Frank$^1$}
\affiliation{$^1$Department of Physiology,
Keck Center for Integrative Neuroscience,
University of California, San Francisco, USA,
e-mail: yura@phy.ucsf.edu}
\affiliation{$^2$ School of Computing, University of Leeds, UK; 
e-mail: a.g.cohn@leeds.ac.uk}

\begin{abstract}
  We discuss the task of reconstructing the topological map of an
  environment based on the sequences of locations visited by a mobile
  agent -- this occurs in systems neuroscience, where one runs into the task of
  reconstructing the global topological map of the environment based
  on activation patterns of the place coding cells in hippocampus area
  of the brain.  A similar task appears in the context of establishing
  wifi connectivity maps.
 
\end{abstract}
\maketitle              

\section{Introduction}
This paper\footnote{Thanks are due to our colleagues and the anonymous referees for their
useful comments.} considers how to infer a topological representation of an
environment from a trace of {\em place information}. By {\em place
information} we assume that this takes the form of a finite set of
{\em places names}, of unknown location, and a time series specifying
when a mobile agent is at these locations. The places may overlap
spatially (and thus more than one place name may be simultaneously
``active''), but are assumed to be unique (not spatially identical).

One example of such data (which was our original motivation) are the
firing traces of {\em place cells} (PCs) of a rat hippocampus after it has
learned a particular environment 
\cite{ODbook}. 
After a period of exposure to a particular spatial
environment, cells in the rat
hippocampus are reliably associated with particular regions of space,
{\em place fields} (PFs) such that the PCs ``fire'' (i.e. have a firing
frequency above a certain threshold) when the rat is in the 
PF, and only then\footnote{Predictive firings have been reported,
  but we ignore that complexity here.}. 
Can the spatial layout of
the environment be recovered purely by inspection of the firing
activity of the PCs?  Recovering a topological map from
PC data in this way has been called the ``space
reconstructing thought experiment'' (SRE)\cite{archive}.  This paper
builds on this unpublished report \cite{archive} where the relevant
neuroscience literature is surveyed in detail. \cite{archive} also suggests the
use of homology theory and mereotopology to analyse the PC data. Here
we explore the second suggestion. 
This problem may occur in other domains; one example is the set
of visible SSIDs\footnote{In the case of multiple base stations with
  the same SSID, the mac address can be used to distinguish them. 
} of wifi base stations\cite{wifiref1,wifiref2}.

\begin{figure}[tbp]
\begin{center}
\includegraphics{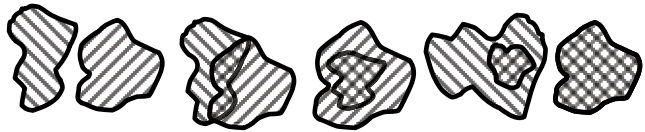}
\end{center}
\caption{The RCC5 relations, from the left: {\sf DC, EC, PO, PP, PPi, EQ}.}
\label{rcc5}
\end{figure}

We take the position here that places are regions rather than points,
not only because it may be hard to determine when the agent is at the
central ``focus'' of a place, but also because this naturally fits
with the two domains mentioned above. This naturally suggests the use of
a region based representation such as RCC\cite{CohnRenz}.
In both these domains it is unlikely that regions will ever
be tangentially connected (the {\sf EC} or {\sf TPP} relations of the
RCC  mereotopological calculus, or similar
relations in other calculi \cite{CohnRenz}). Thus we will
use a purely mereological
calculus, RCC-5\cite{CohnRenz}, illustrated in fig. \ref{rcc5}.
Connectivity between places implies that two regions at least
partially overlap ({\sf PO}). 
The RCC relations can all be defined in terms of the connection primitive {\sf C}($x,y$)
\cite{CohnRenz}, ``$x$ is connected to $y$, as can
a predicate {\sf Con}($x$), 
which is true when $x$ is a simply
connected region (i.e. is one-piece), and a predicate {\sf P}($x,y$)
which is true when $x$ is a part of $y$ (i.e. either {\sf EQ}($x,y$)
or {\sf PP}($x,y$) holds).
The underlying question we are trying to answer here, is whether and
how can a topological map be extracted from the sparse time
series data indicated above. In particular we deliberately assume that
there is no information as to the agent's actual metrical movements,
orientation, odometry or heading. This might be because such
information is hard to obtain or compute, but seems to be an
interesting challenge in its own right.
\label{related}

The idea of computing a topological representation of the environment
of a mobile agent is not new, and has been the subject of
active research in robotics. 
The work on the
``spatial semantic hierarchy''  is perhaps the most relevant here
\cite{Remolina}. This framework provides a
multilevel approach to computing a topological representation, taking
account of metrical, geometrical information and potentially a wide
variety of sensor readings.  Topological information is thus inferred
after several layers of processing, and ``describes the environment as
a collection of places, paths and regions, linked by topological
relations, such as connectivity, order, boundary and
containment''. Places are always points, paths are one dimensional,
and regions two dimensional, e.g. defined by a closed
loop of paths, or by abstraction of a group of places.  The process of
inferring the topology of these entities is by abduction using
circumscription: the underlying idea is to abduce a
minimal topological description which explains the underlying sensor
data. 
Although relevant to the considerations in this paper, the spatial
semantic hierarchy framework differs in several respects; in
particular it assumes, and makes use of non topological information
and it allows for the possibility of aliasing (i.e. that place names
are not necessarily unique). Finally, we mention the Ratslam approach
to simultaneous location and mapping (SLAM) inspired by the neuroscience work
on the rat hippocampus \cite{ratslam3}. This
uses vision, odometry and neural networks to produce a topological map based on
PCs and is contrasted with the more usual approach to SLAM using a
probabilistic approach based on particles.
Ratslam uses both head direction and
PCs (i.e. both directional and positional information) to produce
its maps. Here we only use PCs in order to
concentrate on the purely topological aspects of the problem.

\section{Computing Connectivity}
\label{computingC} 
We assume that we have a finite set of $k$ places, $p_i$,
distinguished by a predicate {\sf Place}($p_i$),  whose intended
interpretations are regions of 2D space.
We also assume a finite set of $n$ times, with a primitive
ordering relation $t_1 < t_2$ which specifies when $t_1$ is temporally
before $t_2$ and a predicate {\sf At}($t,p$) which specifies whether
the agent is at place $p$ at time $t$. For the present, we also assume
that time is sufficiently fine grained such that no place transitions
are missed, i.e. not recorded. We will return to this assumption below
in \S\ref{partial}.
If the agent is at two places simultaneously then we can infer that they
are connected:

$\forall x \forall y \exists t [{\sf At}(t,x) \wedge
 {\sf At}(t,y)] \rightarrow {\sf C}(x,y) $

\noindent
Note that for the converse to hold, we would have to assume that the
agent had made a complete exploration of the environment, in the sense
that every actually physically connected pairs of places had actually
been visited at consecutive times. For the case of rats running
experimental mazes in laboratories, this is certainly the case after a
relatively short period of time. For large scale geographic
environments this may not hold, though it may be a reasonable ``closed
world'' assumption to make, until contrary evidence comes in,
requiring a {\em non monotonic} revision of the topological map. Of
course map revisions may have to be made in any case as a result of
structural changes in the environment or the number or liveliness of
the place indicators.

The set of pairwise place connections can clearly can be computed in
at most $k^2*n$ time, since it simply requires that each time point is
scanned in turn, checking all pairs of places for whether they are
simultaneously active.

In some domains, determining whether the agent is currently at a
particular place may be more problematic. E.g. in the case of
hippocampal PFs, a certain threshold of firing frequency is
required, or in the wifi domain a minimum signal level might be
required. Such additional constraints would need to be factored into
the computation of {\sf At}. This might be a global threshold, or it
may be a local (spatial) threshold to the particular place, or a
temporal threshold (i.e. different thresholds might be applicable at
different times, perhaps due to changing environmental conditions). We
will not consider this further in this paper, though we note that
calculi for reasoning about regions with indeterminate boundaries,
such as the ``egg-yolk'' calculus\cite{CohnRenz} may be relevant.

From the predicate {\sf C}, it is straightforward to build a
connectivity graph, specifying which regions/places are connected
through the overlap relation, and thus the connection relation
{\sf C}($x,y$) holds between them. From the point of view of having a
representation from which it is possible  e.g. to do path
planning, this is all that is
needed. But it would be useful to be able to determine the overall
``topological shape'' of the environment as exposed by the time series
data. Below we show how this might be done; we are not
proposing the particular definitions and concepts below as final, but
rather as illustrative of the kind of analysis possible.

\begin{figure}[tbp]
\begin{center}
\includegraphics{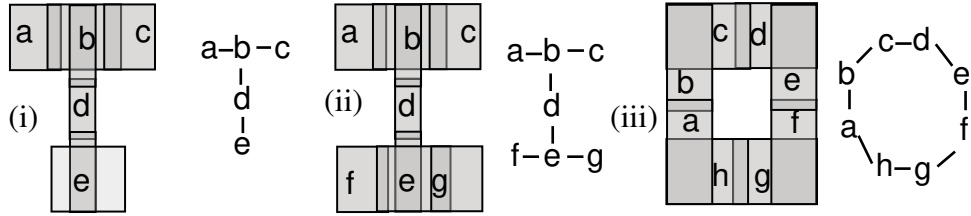}
\end{center}
\caption{ Three 1D environments formed by overlapping places -- the
  darker areas show overlap between regions;
  to the right of each shape is the connection structure constructed by the
  overlaps between places, which
can be distinguished by virtue of their connectivity
structure.}
\label{simpleshapes}
\end{figure}

We first consider environments which are essentially linear, such as
the typical mazes run by rodents.
These might be simple linear runs, or with junctions (e.g. in the
shape of a {\sf T, W, H}),or more complex mazes, with loops (e.g. {\sf
  O, 8, 9}).  In a strictly topological sense, the first set of
examples above are all topologically identical -- they can all be
shrunk to a point.  However, by subdividing the shapes into parts
(corresponding to the simple linear stretches), and then specifying
the connectivity between these parts, these shapes can all be
distinguished -- see fig. \ref{simpleshapes}\footnote{Note that this
  figure, and other similar ones later in the paper are intended to be
  purely illustrative, rather than realistic configurations of actual
  PFs recorded from real rats, or from wifi recordings.}, where
three shapes and their connection structure are depicted. Shapes (i)
and (ii) are topologically identical in the sense that they can both
be shrunk to a point, but the connection structure of their places is
different.  The connection structure can also of course be given
purely symbolically; e.g. (i) is: {\sf C}(a,b), {\sf C}(b,c), {\sf
  C}(b,d), {\sf C}(d,e), and {\sf DC}($\alpha,\beta$), for all other
pairs of regions $\alpha,\beta$. Visualizing this purely symbolic
structure is an interesting problem which we return to below.
In order to formally analyse such linear structures we can make the
following definitions, which group together
collections of regions to form higher level abstractions\footnote{$\exists!^n \alpha
  \Phi(\alpha)$ is syntactic sugar for ``there exist exactly $n$
  $\alpha$ s.t. $\Phi(\alpha)$''. $\exists^n \alpha
  \Phi(\alpha)$ is syntactic sugar for ``there exist at least $n$
  $\alpha$ s.t. $\Phi(\alpha)$''.}.

\noindent ${\sf LinearSegment}(x) \equiv_{def}
{\sf Con}(x) \wedge {\sf SumOfPlaces}(x) \wedge$\\
\hspace*{0.2cm}$\forall y [[{\sf Place}(y) \wedge {\sf P}(y,x)] \rightarrow
          [{\sf End}(y) \vee {\sf Middle}(y) \vee {\sf Junction}(y)]] \wedge$
\\
\hspace*{0.2cm}$\exists!^2 y [{\sf Place}(y) \wedge {\sf P}(y,x) \wedge 
                [{\sf End}(y) \vee  {\sf Junction}(y)]]
$

\noindent ${\sf End}(x) \equiv_{def} 
{\sf Place}(x) \wedge \exists!^1 y [
{\sf Place}(y)  \wedge {\sf C}(x,y) \wedge  \neg {\sf EQ}(y,x) ]
$

\noindent ${\sf Middle}(x) \equiv_{def} 
{\sf Place}(x) \wedge \exists!^2 y [
{\sf Place}(y) \wedge  {\sf C}(x,y) \wedge  \neg {\sf EQ}(y,x)]
$

\noindent ${\sf Junction}(x) \equiv_{def} 
{\sf Place}(x) \wedge \exists^3 y [
{\sf Place}(y) \wedge  {\sf C}(x,y) \wedge  \neg {\sf EQ}(y,x)]
$

\noindent
$ 
 {\sf SumOfPlaces}(x) \equiv_{def}$\\
\hspace*{0.2cm}$\neg \exists y [{\sf P}(y,x) \wedge \forall z  [[
{\sf Place}(z) \wedge{\sf P}(z,x)] \rightarrow  {\sf DC}(y,z)]]
$

Thus {\sf End}s are places only connected to one other place, {\sf
Middle}s only to two other places, and {\sf Junction}s are connected
to three or more other places; {\sf LinearSegment}s are composed of
two places being either {\sf End}s or {\sf Junction}s and all other
places in the  {\sf LinearSegment} are {\sf Middle}s. The predicate 
{\sf SumOfPlaces}($x$) ensures that $x$ is a region every part of
which is part of some place, so that there are no ``extra bits''
of space which are part of $x$ but not part of some place.

\begin{figure}[tbp]
\begin{center}
\includegraphics{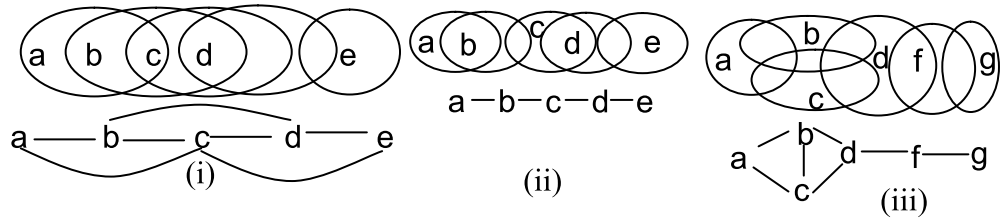}
\end{center}
\caption{A linear track composed of (i) doubly overlapping places, \&
  (ii) singly overlapping places; beneath are  the connection
  structures formed by the overlaps. (iii): An environment with two
  maximal induced paths: abdfg \& acdfg. }
\label{doublePO}
\label{multipleIPs}
\end{figure}

These definitions achieve the desired effect in linear environments
providing that no places are long enough to overlap more than one
place in each direction (i.e as in fig. \ref{doublePO}(ii)). E.g.,
consider fig. \ref{doublePO}(i) -- the layout is still
clearly in some sense linear, but as the connection graph shows, it is
not so in any very straightforward way. 

One approach to this might be
to use the idea of an {\em induced path}
-- i.e. a
sequence of nodes in a graph such that each is connected to two
neighbours in the path, except for the two end nodes, and with the
proviso that there are no ``short cuts'' (i.e. direct links) between
any two nodes in the path using other edges in the graph (which would
be indicated by the presence of a 3-clique)\footnote{In the definition
  below, we also rule out trivial induced paths of length one
  (i.e. with just two nodes and one edge).}. Thus in the
graph in fig. \ref{doublePO}(i), abde, ace, bce, acd,  are all (maximal)
induced paths. However this notion is not entirely satisfactory
since it does not define a unique path for the environment (since there
are two induced paths from a to e). In this case we can note that
there is a unique longest induced path (abcd); of the others, ace has
the same start and end nodes, and the other two each have an end in
common with the longest. None are disjoint.  We also note that all the
nodes in the non longest induced paths are within one edge of a node
on the longest induced path (i.e. c is directly linked to a,b,d and
e). It is not guaranteed that there is a unique longest induced path
-- either because there are two entirely disjoint such paths, or
because there are alternatives with common nodes -- e.g. see fig.
\ref{multipleIPs}(iii). We thus propose the following definition for what we will call
{\em quasi linear} segments.
\begin{figure}[tbp]
\begin{center}
\includegraphics{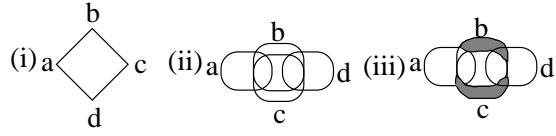}
\end{center}
\caption{A configuration in which by considering  place
  trails from than two places simultaneously, a refined map
  can be obtained.}
\label{mereofig}
\end{figure}

\noindent${\sf QuasiLin}(x) \equiv_{def}$\\
\hspace*{0.2cm}$\exists w [{\sf P}(w,x) \wedge {\sf InducedPath}(w) \wedge {\sf
  SumofPlaces}(x) \wedge$\\
\hspace*{0.4cm}$\forall y [[{\sf P}(y,x-w) \wedge {\sf Place}(y) ] \rightarrow {\sf C}(y,w)]]
$

\noindent$ {\sf InducedPath}(w) \equiv_{def}
$\\
\hspace*{0.2cm}${\sf SumOfPlaces}(w) \wedge {\sf Con}(w) \wedge 
\neg \exists v [{\sf 3Clique}(v) \wedge {\sf P}(v,w)] \wedge$\\
\hspace*{0.2cm}$\exists (s,x,y,z) [{\sf Place}(x) \wedge 
{\sf Place}(y) \wedge \neg {\sf EQ}(x,y) \wedge  {\sf Place}(s) \wedge {\sf P}(s,z) \wedge
$ \\
\hspace*{0.2cm}$ {\sf EQ}(w,x+y+z) \wedge 
\forall u [[{\sf Place}(u) \wedge {\sf P}(u,z)] \rightarrow \neg {\sf Con}(w-u)]]
$

\noindent$ {\sf 3Clique}(w) \equiv_{def} \exists (x,y,z) [ $\\ \hspace*{0.2cm}${\sf Place}(x) \wedge
{\sf Place}(y) \wedge
{\sf Place}(z) \wedge {\sf EQ}(w,x+y+z) \wedge \neg {\sf EQ}(x,y) \wedge
$\\
\hspace*{0.4cm}$
\neg {\sf EQ}(x,z) \wedge
\neg {\sf EQ}(y,z) \wedge
{\sf C}(x,y) \wedge
{\sf C}(x,z) \wedge
{\sf C}(z,y)]
$

Returning to fig. \ref{doublePO}(i), ${\sf QuasiLin}$(a+b+c+d+e) is
true since it has at least one induced path in it, and all the other
places in it are directly connected to a place in that induced
path. Exactly what definition of ``linearity'' might be appropriate in
a particular domain
will depend on what is required or suitable for the application.
E.g., ${\sf QuasiLin}(x)$ will be true if $x$ is the union of a
simple path of length $m$ and a clique of size $n$ s.t. exactly two
nodes are in common between the clique and the simple path. It would
be straightforward to eliminate this case, or to allow only cliques of
up to a size $m$ to occur within a quaslinear path.

Another important grouping of places are ``open spaces'' which are
likely to be identified by clusters of many places, though not
necessarily in the form of a clique (though they may well contain
cliques). Having identified cliques, and linear segments, a natural
step would be to replace these by ``super nodes'', and then continue
analysing the environment at this more abstract level. Indeed, there
are already existing approaches to analysing and drawing graphs which
take this approach, e.g. \cite{muller,Friedrich}. E.g., ``open
spaces'' may frequently be represented as cliques of cliques.

One other aspect of connectivity analysis not so far mentioned
explicitly is determining whether there are cycles in the environment
(caused by circular structures in a linear environment, or by
obstacles in an open environment). A graph theoretic approach to this
would be to look for {\em chordless cycles}, i.e. circular induced
paths of length at least four. As a trivial example of this, consider
fig. \ref{simpleshapes}(ii); in practice this approach is likely to require
refinement to properly capture the required notion.

\section{Variants of the mapping task}
\label{variants}
In this section we consider a number of variants of the basic mapping
task and how they might be achieved.

\label{mereo}

\begin{sloppypar}
\noindent {\bf Computing Connectivity Mereologically}:
The approach outlined above shows how we can compute connection
information from knowledge that the agent is simultaneously {\sf At}
two places. This gives rise to binary connectedness information, and
thence (in RCC-5) the knowledge that particular {\em pairs} of places
partially overlap ({\sf PO}). It is not possible from the connection
structure though to infer that any triple (quadruple...) of places
overlap, even though this information might in fact be readily
apparent in the place trail (indicated by an agent being
simultaneously at three (four...) places). We could therefore compute
a more fine grained representation, in which we explicitly represent
those intersections of places which are known to exist, and similarly
those relative complements of places known to exist.  E.g. from the
place trail: {\sf At}(t1,a), {\sf At}(t2,a), {\sf At}(t2,b), {\sf
  At}(t3,a), {\sf At}(t3,b), {\sf At}(t3,c), {\sf At}(t4,b), {\sf
  At}(t4,c), {\sf At}(t5,b), {\sf At}(t5,c), {\sf At}(t5,d), {\sf
  At}(t6,d), using pairwise connections, as in \S\ref{computingC},
then a connection structure as in fig. \ref{mereofig}(i) would be
computed. Not knowing about which subregions can actually exist, it
would be reasonable to produce a map such as in fig.
\ref{mereofig}(ii). However, by inspecting the above place trail, we
can infer that the regions a--b--c--d, a+b--c--d, a+b+c--d, b+c-a--b,
b+c+d--a, and d--a--b--c all exist, but there is no evidence to
support the existence of any of the other five Boolean combinations of
the four places, a, b, c, d which exist in fig. \ref{mereofig}(ii).
Thus we can build a simplified map as in fig. \ref{mereofig}(iii),
in which these regions do not exist (indicated by shading).
\end{sloppypar}

\noindent {\bf Partial information}:
\label{partial}
It is believed that PCs form a cover for the environment the
rat has explored, i.e. wherever the rat is, at least one PC
will be firing.  However for the SRE,  an
external observer is receiving signals from a set of electrodes, and
only a subset of the PCs will actually be recorded and thus only
partial information about the set of places active at any
time. At some times, there may be no active PCs
being recorded, and thus there will be ``temporal gaps''. The question
is, what can we say about the nature of the environment given only
such partial information? In the wifi domain, presumably
full information would always be available; there might still be
temporal gaps, because no base station is in range, but that is
different to not being able to detect a base station which is in
range. One might want to regard the union of all locations where there
is no base station in range as a $k+1$st place.  The discussion
below concerns domains where there is only partial information.
First we define the notion of a temporal gap:

\noindent ${\sf Gap}(t_1,t_2) \equiv_{def}
\exists  (x_1,x_2) [{\sf At}(t_1,x_1) \wedge
 {\sf At}(t_2,x_2) \wedge$\\
\hspace*{0.2cm}$\forall (t_3,x_3) [[ t_1 < t_3 \wedge t_3 < t_2] \rightarrow \neg {\sf
  At}(t_3,x_3)]
$

\noindent We can now write a rule which allows us to infer that the agent must
be at at least one place during a gap, and that these places form a
connected region which is itself connected to all the places where the
agent is at at $t_1$ and $t_2$.

\noindent $\forall (t_1,t_2,x_1,x_2)  [[{\sf Gap}(t_1,t_2)
\wedge {\sf At}(t_1,x_1) \wedge {\sf At}(t_2,x_2)]
 \rightarrow\\
\indent
\exists x_3 [{\sf SumOfPlaces}(x_3) \wedge {\sf C}(x_1,x_3) \wedge
 {\sf C}(x_2,x_3) \wedge {\sf Con}(x_3)]]
$

There might be more than one path $x_3$ linking the places at
$t_1$ and $t_2$, however, we cannot infer this without
evidence (such as metric
information about speed and distance travelled, but we do not consider
such possibilities here).

There is (at least) a second way in which the underlying place
information might be partial. We made the explicit assumption (in
\S\ref{computingC}) that time is sufficiently fine grained that
no place transitions are missed. If this is not the case (because the
speed of the agent is fast with respect to the recording granularity),
then gaps may occur even if there are no missing place sensors. In this case
we would need a modified version of the rule above, since it may be
that no new places need to be inferred to fill in the gap, but rather
that at least one of the places at time $t_1$ directly connects with
at least one of the places at time $t_2$.
Some form of non monotonic reasoning is likely to be needed in general
for reasoning in the presence of such kinds of partial knowledge, in
order to perform a domain closure, or to minimize the number of places
assumed to exist (as in the spatial semantic hierarchy discussed in
\S\ref{related}).

\section{Final Comments}
We have discussed the problem of computing topological maps from
knowledge of place trails, a task applicable at least in two
identified domains. There are many ways in which this work could be
extended.  E.g. we could consider how to turn the symbolic topological
representations into a graphic visualisation\footnote{Note that we are
  interested in a visualisation in which the regions are explicitly
  represented as regions; thus although there are standard techniques
  for laying out planar graphs (such as the connection graphs above)
  with a reasonably balanced vertex distribution and straight line
  edges, which could be buffered to produce regions, this makes the
  connections into regions rather than the regions themselves.}
automatically. Of course any such depiction will inevitably have
metric qualities, but these must be ignored when interpreting the
visualization\footnote{In fact, in both the domains we are
  considering, there are some very approximate metric qualities which
  could be inferred, since PFs are known to have a typical size (5cm x
  5cm to about 35cm x 35 cm) and wifi base stations similarly have a
  maximum range.}. Any qualitative spatial description will always
have many metric realizations. One approach is to
diagrammatic reasoning techniques (e.g. \cite{howse2}) on
visualising Euler diagrams.

We have already conducted some
experimental work with artificial data and are currently collecting
real data and will then evaluate the ideas sketched here, and refine
them as appropriate. We may also consider other variants of the
problem, e.g. scenarios with multiple agents (where the {\sf
  At} predicate has a third argument
indicating the agent).

\vspace*{-0.1in}

\end{document}